# Reconfigurable, Temperature Resilient Phase-Change Metasurfaces Fabricated via High Throughput Nanoimprinting Lithography

*Carlota Ruiz de Galarreta+\*[1,3] Yinghao Zhao+[1,2], Jose Mendoza-Carreño[1], José M Caicedo[4], José Santiso[4], C David Wright[3], M Isabel Alonso[1], Agustin Mihi\*[1]*

[1] Institute of Materials Science of Barcelona ICMAB-CSIC; Campus UAB, Bellaterra, Spain
[2] School of Physics, Beijing Institute of Technology, Beijing 100081, China
[3] Centre for Metamaterials Research & Innovation, University of Exeter, Exeter EX44QF, UK
[4] Catalan Institute of Nanoscience and Nanotechnology (ICN2), Campus Universitat Autonoma de Barcelona, Bellaterra, Catalonia, Spain

E-mail: *cruiz@icmab.es, *amihi@icmab.es

+ Y.Z and C.RdG contributed equally

**Abstract**

The combination of metasurfaces with chalcogenide phase-change materials is a highly promising route towards the development of multifunctional and reconfigurable nanophotonic devices. However, their transition into real-world devices is hindered by several technological challenges. This includes, amongst others, the lack of large area photonic architectures produced via scalable nanofabrication methods, as required for free-space photonic applications, along with the ability to withstand the high temperatures required for the phase-change process. In this work, we present a scalable nanofabrication strategy for the production of reconfigurable metasurfaces based on high-throughput, large-area nanoimprint lithography that is fully compatible with chalcogenide phase-change materials processing. Our approach involves the direct imprinting of high melting point, thermally stable $TiO_2$ nanoparticle pastes, followed by the deposition of an $Sb_2Se_3$ thin film as the phase-change material active layer. The patterned titania film enables the creation of thermally robust metasurfaces, overcoming the limitations of conventional polymer-based nanoimprint approaches. The versatility of our approach is showcased by producing phase-change devices with two distinct functionalities: (*i*) metasurfaces with tunable spectral band switching and amplitude modulation capabilities across the near- to mid-infrared, and (*ii*) reconfigurable chiral metasurfaces, whose chiroptical activity can be switched between the visible and the near-infrared. Experimental results show excellent agreement with numerical simulations and reveal high uniformity across large areas. This work provides a universal, thermally robust and scalable platform for the production of reconfigurable metasurfaces based on phase-change materials, paving the way to low-cost, photonic devices with reconfigurable optical responses that could be extended far beyond the applications demonstrated here.

# 1. Introduction

As the field of nanophotonics keeps expanding and maturing, research efforts have turned to the development of active and reconfigurable devices, whose optical behavior can be altered or modified post-fabrication via the incorporation of materials capable of reversibly responding to external physical stimuli.[1,2] Among the various strategies available to yield reconfigurable nanophotonic devices, the use of chalcogenide phase-change materials (PCMs) has emerged as one of the strongest approaches due to their inherent low power consumption, fast switching speeds, and abrupt refractive index contrast between their fully reversible amorphous and crystalline phases.[3–6] Traditionally employed in commercial non-volatile optical and electrical memories,[4] chalcogenide PCMs such as the archetypal alloy $Ge_2Sb_2Te_5$ (GST) can reversibly transition between amorphous, crystalline, and intermediate states through thermal stimuli. Such stimuli are typically provided by time and power controlled electrical or optical pulses, supplied, in the case of nanophotonic devices, by resistive heaters, or laser excitation respectively.[3–5,7,8] In particular, crystallisation is achieved by heating the material above its crystallisation temperature, while amorphisation involves a melting step, followed up by a rapid cooling of the molten phase (up to tens of °C per nanosecond for GST) to prevent recrystallisation.[3,4] As a result of their remarkable physical properties, PCMs have over the past decade PCMs been successfully combined with nanophotonic architectures such as metasurfaces, integrated photonic circuits, or thin film multilayer stacks.[2,3,9] This has led to the development of devices with diverse functionalities including non-mechanical beam steering,[8,10] reconfigurable metalensing,[11–13] switchable image pre-processors,[14,15] amplitude modulation,[16,17] tunable structural colouring and filtering,[18,19] or neuromorphic photonic computing to name but a few.[20,21]

Although significant advances in the field have been made in recent years, the successful integration of PCMs into reconfigurable metasurfaces for real-world devices remains a formidable challenge due to several engineering crossroads arising from such a technological combination,[2,3] including thermal and chemical incompatibilities, or device endurance to name a few.[22] Another critical –and often overlooked— aspect of such technology is the combination of PCMs with large scale, cost efficient, and large area nanofabrication technologies.[2] To date, PCM based metasurfaces have been mostly fabricated using electron beam lithography combined with reactive ion etching and/or lift-off techniques[3,7,8,10,13,14,17]: processes that offer excellent versatility, pattern fidelity and resolution, but at the same time suffering from low throughput, limited scalability, and high cost. Alternative microfabrication techniques based on direct laser writing of crystalline meta-atom arrays on amorphous PCM films have been recently explored and experimentally validated in the mid- to far-infrared.[23,24] However, although highly scalable, this approach is limited in terms of design versatility and optical performance; it also suffers from reduced resolution arising from the diffraction limit and optical penetration depth of PCMs, making it not well suited to the production metasurfaces operating in the near-infrared and visible spectral range. In this context, nanoimprint lithography (NIL) offers a compelling alternative due to its simplified implementation, low cost, high resolution, and suitability for large-area patterning.[25,26] Nevertheless, and in spite of its excellent scalability potential, to date the use of NIL towards large scale production of reconfigurable phase-change metasurfaces has remained unexplored. In this work, we address this key omission by introducing a high-throughput, low-cost NIL nanofabrication route suitable for patterning PCM based reconfigurable metasurfaces over large areas. Our approach enables the direct nanoimprinting of mesoporous TiO₂ (m-TiO₂) paste made of anatase nanoparticles, for subsequent deposition of Sb₂Se₃ thin films (Figure 1a). The Sb₂Se₃ coating provides a high refractive index tunable environment through its amorphous and crystalline phase transitions (Figure 1b). On the other hand, the nanoimprinted m-TiO₂ backbone not only offers excellent resolution and pattern homogeneity over large areas, but crucially –and contrary to conventional polymer-based resins— it grants high thermal stability capable of withstanding temperatures of at least 1000 °C for 1H.[27] This aspect is critical from a device functionality point of view, as temperatures required for a successful cycling of chalcogenide PCMs can go above 170 °C and 600 °C for crystallisation and re-amorphisation processes respectively.[3,4] To illustrate the potential and versatility of our approach, we have developed a set of $Sb_2Se_3$-based reconfigurable metasurfaces, showcasing two distinct functionalities. As depicted in Figure 1c, the first type of metasurfaces consists of square arrays of microholes supporting high quality factor resonances, whose resonant spectral position can be switched by changing the phase of the $Sb_2Se_3$ layer between amorphous, crystalline, and intermediate states. This design provides reconfigurable filtering and modulation across the near and mid- infrared, with experimental absolute modulation depths as high as 60% in both reflection and transmission. The second type of metasurface studied (Figure 1d), is constituted of a hexagonal array of triskelion motifs engraved in the m-$TiO_2$ and coated with $Sb_2Se_3$. This chiral array exhibits a strong resonant chiroptical response that can be switched between the visible and the near-infrared when the $Sb_2Se_3$ layer transitions from amorphous to crystalline. All the devices developed and optically characterised showed good agreement with numerical simulations, confirming the effectiveness and versatility of our approach. Therefore, we believe our approach could be generalized towards the development of scalable and low-cost phase-change metasurfaces with high temperature resilience, for additional functionalities to those shown here including, but not limited to, active beam steering, dynamic colour displays, reconfigurable optical analog computing and more

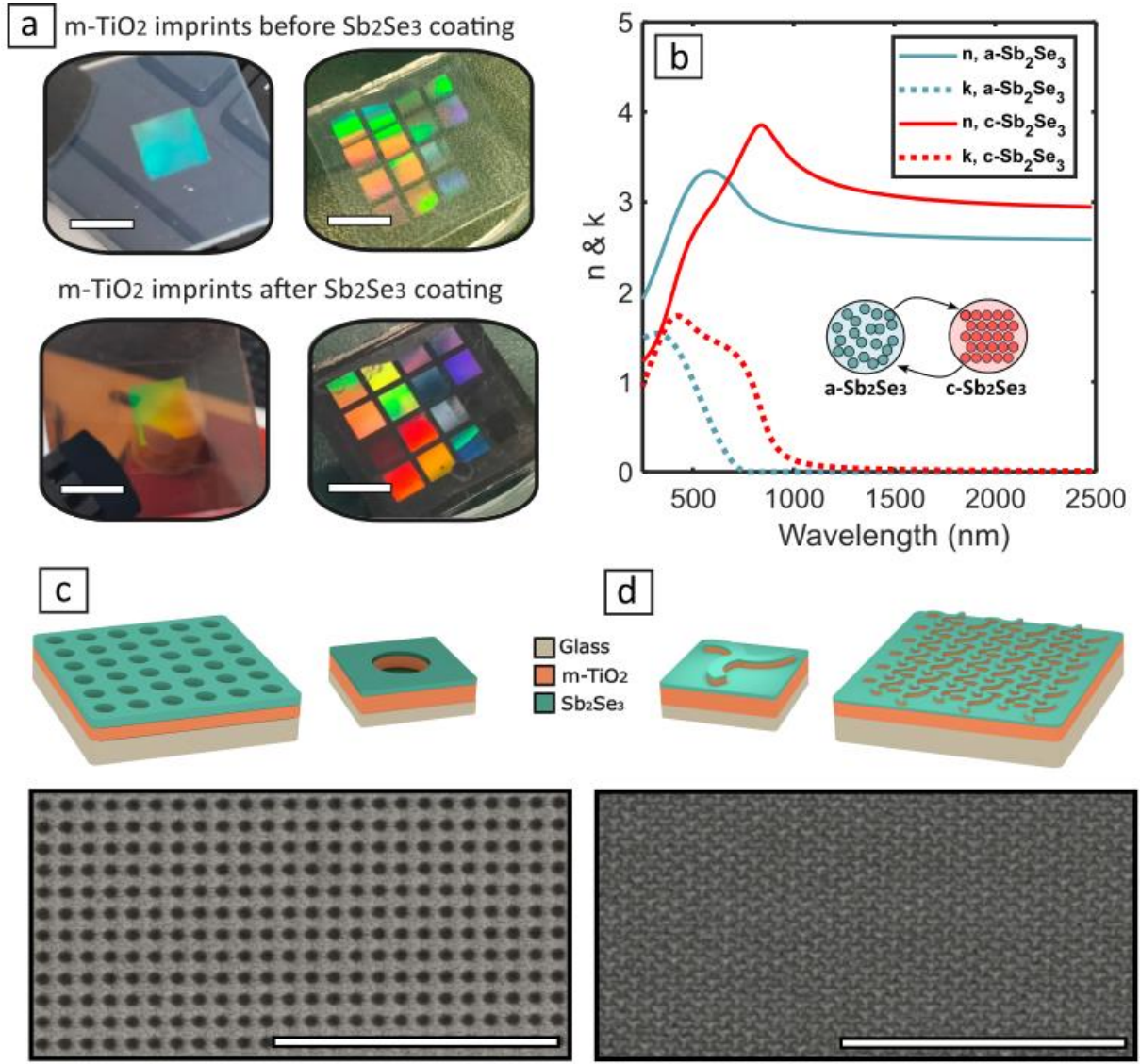

**Figure 1. Schematics, optical properties, and morphology of large-scale nanoimprinted phase-change metasurfaces** (a) Optical images of m-TiO₂ imprinted arrays before (top) and after (bottom) deposition of Sb₂Se₃ (Scale bars are 1 cm). (b) Refractive index (*n*, solid lines) and extinction coefficient (*k*, dashed lines) of our amorphous (a-Sb₂Se₃, blue) and crystalline (c-Sb₂Se₃, red) films, obtained by spectroscopic ellipsometry. (c) Schematic illustration (top) and SEM image (bottom) of ordered hole m-TiO₂ arrays coated with Sb₂Se₃. (d) Schematic illustration (top) and SEM image (bottom) of hexagonal m-TiO₂ triskelia arrays coated with Sb₂Se₃. Scale bars in SEM images is 10 µm.

## 2. Results and discussion
### 2.1. Fabrication and architecture of nanoimprinted PCM metasurfaces

Figure 2a summarizes the fabrication process undertaken for the metasurfaces showcased in this work. First, a commercial $TiO_2$ paste (used herein as the NIL resist) made of 20 nm diameter anatase nanoparticles was vigorously diluted in ethanol in a 1:3 (w/w) ratio, and then spin coated onto clean glass substrates at 2000 rpm for 10 s, with an acceleration of 1000 rpm.s$^{-1}$. Next, pre-patterned hybrid stamps made of hard-PDMS (h-PDMS) and soft-PDMS (s-PDMS) as the backbone were used to imprint the desired patterns onto the $TiO_2$/ethanol spin coated films on a hot plate at 125 °C for 10 min. The patterned films were then baked at 500 °C for one hour to eliminate the organic fraction of the paste leading to the m-$TiO_2$ arrays. Note we choose an m-$TiO_2$ annealing temperature of T= 500 °C as this is the highest temperature available in our hot plate, but annealing routines at values of up to 1000 °C have also been successfully employed,[27] making the herein fabrication process particularly attractive for systems and/or applications which require to withstand high temperatures. Moreover, around 800 °C anatase nanoparticles transition to rutile,[27,28] providing a higher refractive index medium which potentially offers additional design degrees of freedom to our approach. Following the fabrication flow shown in Figure 2a, after obtaining the m-$TiO_2$ nanostructured backbone, an $Sb_2Se_3$ layer was deposited on top via pulsed laser deposition. We selected $Sb_2Se_3$ as the phase-change material due to its low optical losses in both amorphous and crystalline states across the near- to mid- IR when compared to more traditional GeSbTe ternary alloys.[29–31] A detailed account of the materials, protocols and instruments used can be found at the *Experimental section*

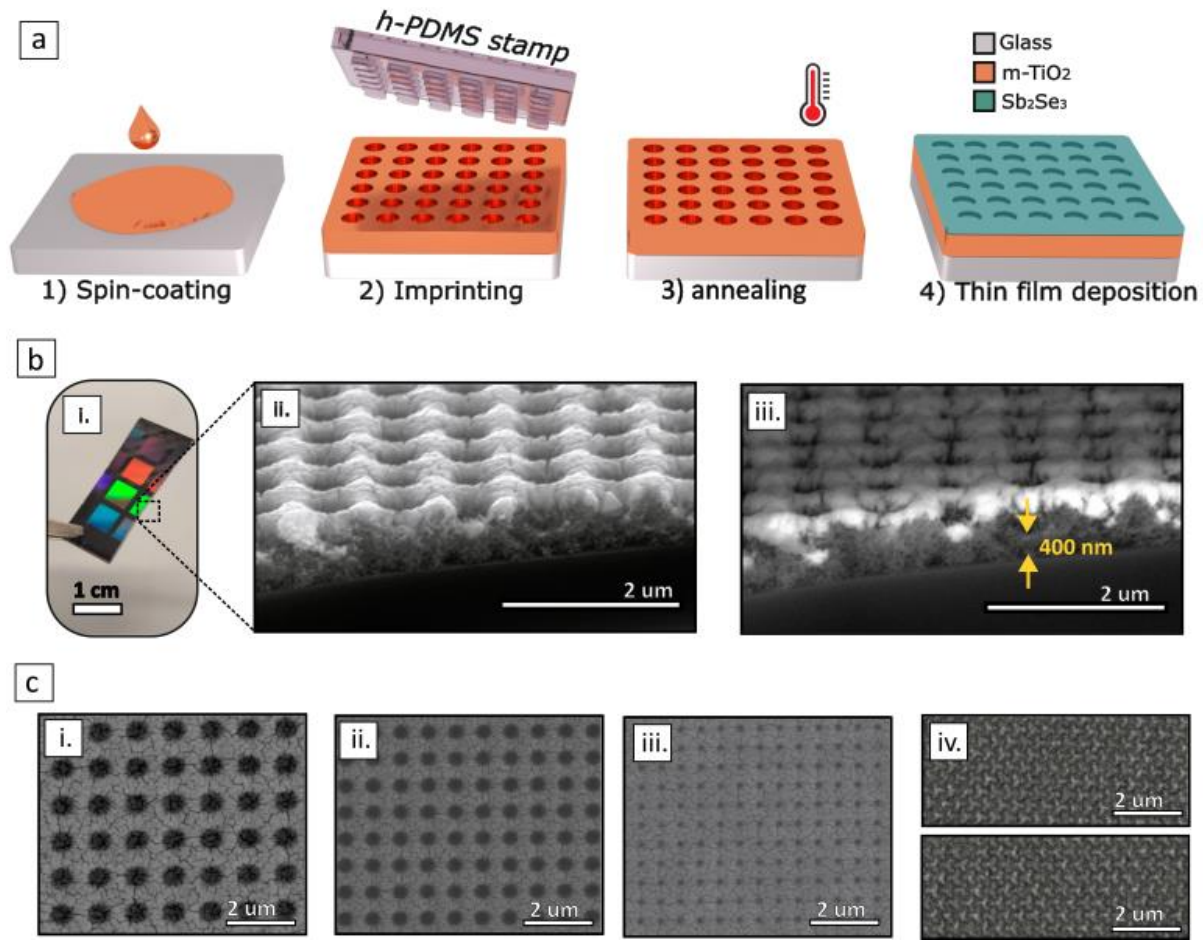

**Figure 2. Fabrication and structural characterisation of nanoimprinted PCM metasurfaces**
(a) Flowchart of the fabrication process summarizing the key steps. (b) Cross sectional device characterisation: (i) Optical image of a patterned chip showing multiple device areas prepared for SEM imaging (ii) Cross-sectional SEM image of a completed device, and (iii) its corresponding back scattering electron detection version, revealing fine details about the porous m-TiO$_2$ layer, sandwiched between the Sb$_2$Se$_3$ coating (brighter region) and the glass substrate (darker region). (c) Top-view SEM images showing a variety of devices fabricated and optically characterised in this work. (i-iii) Square arrays of holes with different lattice parameters and hole diameters. (iv.) Hexagonal arrays of left-handed (top) and right-handed (bottom) triskelion motifs.

The cross-sectional SEM inspection (Figure 2b) of one of the fabricated devices shows the presence of both the PCM coating and the m-TiO2 backbone. In particular, the back scattering electron image shown in Figure 2b(iii) reveals an m-TiO2 residual layer of ~400 nm, which was found to be present in all the devices. Moreover, to verify the stoichiometry of the Sb$_2$Se$_3$ overlayer, we performed energy-dispersive spectroscopy (EDS). Results from such measurements can be found in the supplementary information section S1, where we show atomic percentages of Sb (42.64%) and Se (57.36%). This corresponds to an atomic ratio of **Sb: Se ≈ 2: 3** within the EDS error, indicating that the stoichiometry of the Sb2Se3 target was properly transferred to the film.

Finally, in Figure 2c we present a set of PCM metasurface architectures fabricated and optically characterised in the next sections, elucidating the versatility and geometrical scalability of our fabrication strategy. In particular, in Figures 2c(i-iii) we show several square arrays of microholes with a variety of lattice parameters and hole diameters, which were employed for spectral band switches and modulators across the near to- mid-infrared (in section 2.1). On the other hand, Figure 2c(iv) shows left-handed and right-handed hexagonal arrays of triskelion motifs, whose chiroptical activity can be switched between the visible and infrared regimes, as further discussed in section 2.3.

**2.2. Phase-change scalable metasurfaces for infrared band switching and modulation.**

The first set of PCM metasurfaces studied are intended for infrared reconfigurable band switching and modulation. This functionality has been successfully exploited in different ways and spectral bands for key cutting-edge applications such as active color displays,[32,33] multispectral thermal imaging,[34] or reconfigurable image processing.[14,35] However, to date their fabrication has always relied on expensive, cleanroom-based, conventional electron-beam lithography carried out over small areas, thereby limiting their scalability potential.

A generic unit cell of the proposed device is depicted in Figure 3a. It consists of an m-TiO$_2$ slab lying on a glass substrate, and imprinted with holes of 200 nm depth, forming a square array with an m-TiO$_2$ residual layer of thickness 400 nm. An Sb$_2$Se$_3$ film of thickness $t_{Sb2Se3}$ = 160 nm covers the holes, providing a high refractive index tunable dielectric environment to the device upon transitioning the Sb$_2$Se$_3$ from its amorphous to crystalline states. The Sb$_2$Se$_3$ film is protected with a 15 nm thick Al$_2$O$_3$ layer (not represented for simplicity) to prevent its oxidation.[8,22] These geometrical parameters were kept constant for all the devices, while a set of arrays with different hole diameters (D) and lattice constants (Λ) were fabricated.

Figure 3b displays the measured transmittance spectra of a device with $D$ = 0.64 μm, and $Λ$ = 1.06 μm for both amorphous (a-Sb$_2$Se$_3$, turquoise curve) and crystalline (c-Sb$_2$Se$_3$, red curve) states. Crystallisation of the device was carried out using a hot plate (as described in *section 4.5)*. As can be observed, the metasurface exhibits a set of sharp resonances manifested as transmittance dips across the near- to mid- IR spectral range (see *section 4.6* for details on the optical measurements). These resonances are spectrally shifted by Δλ ~ 200 nm upon crystallisation of the phase-change layer, due to an abrupt increase of its refractive index, accompanied by a minor decrease of the quality factor as a result of the non-negligible absorption coefficient of crystalline Sb$_2$Se$_3$ (as displayed in Figure 1b). Our experimental results are also in good agreement with finite element simulations carried in COMSOL multiphysics, and shown in Figure 3c for comparison (details, material properties and boundary conditions employed in our finite element models can be found in *section 4.4*). Finally, in Figure 3d we show the absolute modulation depth of the device in transmission (MDT), obtained from the experimental values as:[16]

$$MDT(\lambda) = |T_{am}(\lambda) - T_{cr}(\lambda)| \qquad \text{Eq.1}$$

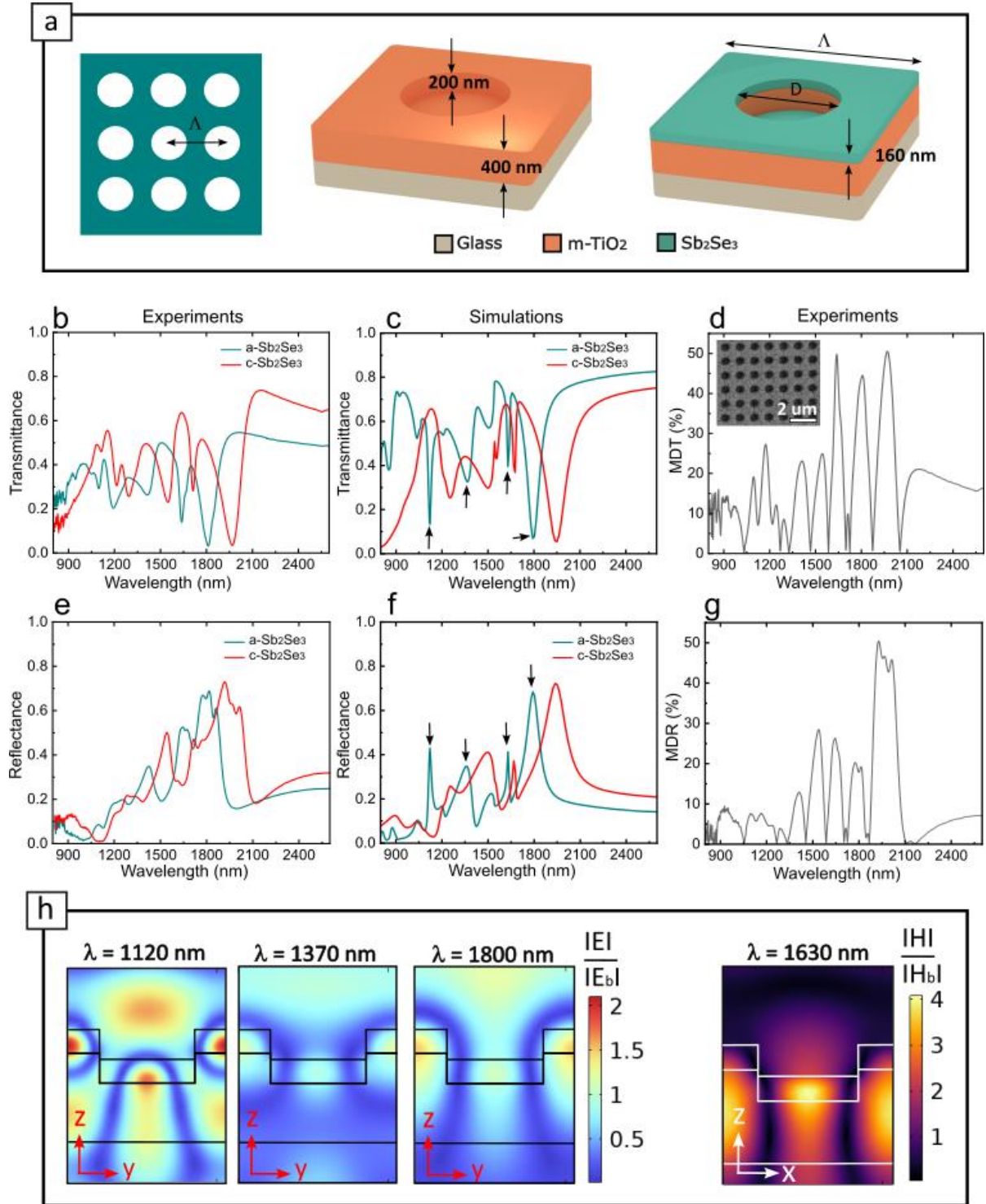

**Figure 3 Nanoimprinted Sb$_2$Se$_3$-based metasurfaces for spectral filtering and modulation.** (a) Schematic and geometrical characteristics of our metasurface unit cell, consisting of a mesoporous TiO$_2$ (m-TiO$_2$) square array of holes, coated with an Sb$_2$Se$_3$ layer. (b-c) Experimental (b) and simulated (c) transmittance spectra for amorphous (a-Sb$_2$Se$_3$, turquoise) and crystalline (c-Sb$_2$Se$_3$, red) states for a device with D = 0.65 μm, and Λ = 1.06 μm (d) Experimentally obtained modulation depth in transmission (MDT), revealing absolute modulation values up to 50% (inset shows an SEM image of the device). (e-f) Experimental (e) and calculated (f) reflectance spectra of the same device for a-Sb$_2$Se$_3$ and c-Sb$_2$Se$_3$ states. (g) Experimentally obtained modulation depth in reflection (MDR) (h) Simulated enhanced field distributions (|E|/|E$_b$| and |H|/|H$_b$| for TE and TM modes respectively) at representative resonant wavelengths identified with black arrows in (c) and (f), for excitation with x-polarised light.

where $T_{am}$ and $T_{cr}$ are the transmittance spectra for the metasurface with amorphous and crystalline $Sb_2Se_3$ respectively. Absolute experimental modulation values as high as 50% can be observed at different sharp spectral bands across the mid IR. Figures 3e and 3f display the measured and simulated reflectance spectra of the same device for both a-$Sb_2Se_3$, and c-$Sb_2Se_3$. As in previous results from transmission measurements, a set of resonances can be identified, here however manifested as complementary sharp reflectance peaks, indicating that the transmission dips observed in Figures 3b and 3d correspond to light being specularly reflected to free-space by the metasurface. Modulation depths achieved in reflection (MDR) were found to be in line with transmission values (i.e. around 50%, as revealed in Figure 3e). To elucidate the effect of the periodic patterning in our films, optical measurements were also carried out outside the patterned region (i.e. at regions with unpatterned multilayer stacks made of glass, 400 nm of m-$TiO_2$, 160 nm of $Sb_2Se_3$ and 15nm of $Al_2O_3$). As shown in the supplementary information *section S2*, these measurements revealed the presence of a single vertical Fabry-Pérot resonance coming from interferences at the $Sb_2Se_3$ and m-$TiO_2$ layers. Therefore, the rich, multiresonant optical behaviour of our metasurfaces arises exclusively from the as-designed geometrical architecture. In particular, the $Sb_2Se_3$ film endows the structure with a high refractive index, generating Mie-like resonators, while the periodic perturbation coming from the $TiO_2$ underlayer adds extra momentum to launch hybrid leaky-guided modes (LGM) inside the m-$TiO_2$ residual layer, the latter behaving as an intermediate refractive index waveguide ($n_{m-TiO2}$ ~1.6) forming an asymmetric system cladded by the lower refractive index glass substrate ($n_{glass}$~1.5). Further insights into the nature of these resonant modes was obtained via analysis of the electromagnetic field enhancement at the four resonant wavelengths highlighted with black arrows in Figures 3c and 3f. The enhanced electromagnetic field distribution under x-polarized light of such modes is depicted in Figure 3h. These were calculated as $|E|/|E_b|$ and $|H|/|H_b|$, where $|E_b|$ and $|H_b|$ are the maxima of the modulus of the electric and magnetic fields, taken from air domains far away from the device). Following a modified Fabry–Pérot model,[36,37] we classify the transverse electric and magnetic modes as TE($m, n, l$) and TM($m, n, l$), respectively, where $m$, $n$ and $l$ represent the mode quantum numbers or number of electric/magnetic antinodes along the $x$, $y$, and $z$ axes. Based on this definition, we find three LGM resonances, which can be classified as TE(1,3,2) at λ= 1120 nm, TE(1,3,1) at λ= 1800 nm, and TM(3,1,1) at λ= 1630 nm. An additional Mie-like electric dipole resonance can be observed at λ= 1370 nm, where electric fields are weakly confined inside the high refractive $Sb_2Se_3$ domains only. This less intense electric field confinement is also manifested as a lower quality factor in the reflectance and transmittance spectra (shown in Figures 3f and 3c) when compared to the other higher Q-factor LGM resonances, whose electric and magnetic field confinements are also visibly higher (see Figure 3h).

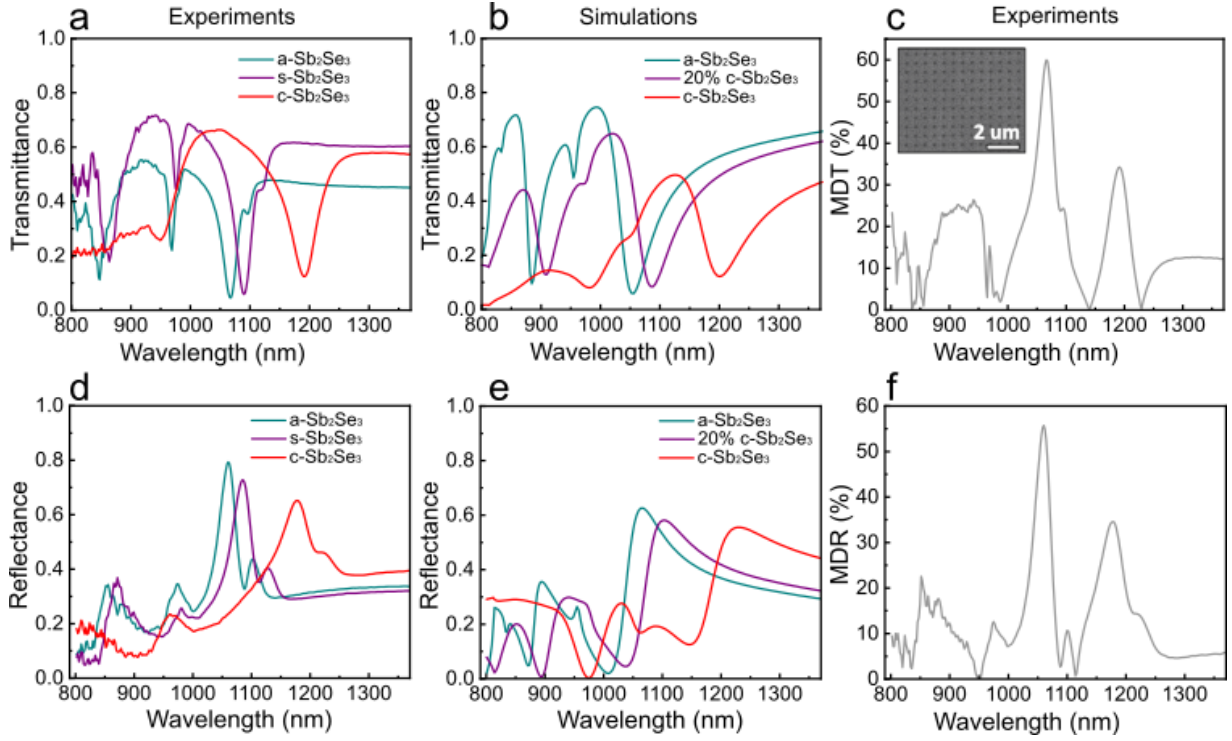

**Figure 4.** (a-b) Experimental (a) and simulated (b) transmittance spectra for amorphous (a-Sb$_2$Se$_3$, turquoise), semicrystalline (s-Sb$_2$Se$_3$, purple) and crystalline (c-Sb$_2$Se$_3$, red) states for a device with D = 0.26 μm, and Λ = 0.56 μm. (c) Experimentally obtained modulation depth in transmission (MDT), revealing absolute modulation values up to 50% (inset shows an SEM image of the device). (d-e) Experimental (d) and calculated (e) reflectance spectra of the same device for a-Sb$_2$Se$_3$, s- Sb$_2$Se$_3$, and c-Sb$_2$Se$_3$ states. (f) Experimentally obtained modulation depth in reflection (MDR).

To explore the spatial resolution achievable with our PCM metasurface fabrication routine, we fabricated and tested devices with significantly smaller geometrical features, such as those shown in Figure 2c(iii) (with a period Λ = 0.56 μm, and a hole diameter D = 0.26 μm). Moreover, multilevel switching capabilities were also investigated via partial crystallisation experiments (described in *section 4.5*). In Figure 4a, we show the measured transmittance spectra for amorphous (turquoise), semicrystalline (purple) and crystalline (red) Sb$_2$Se$_3$. Due to a significant reduction of the metasurface dimensions with respect to the previous device, the multiresonant behaviour is shifted towards the blue, falling now in the near-infrared regime. Our fabrication routine can be therefore employed to customise the position of the LGM resonances across the IR, by simply changing the geometrical parameters of the h-PDMS/s-PDMS hybrid stamps. As it can be observed, a progressive crystallisation of the Sb$_2$Se$_3$ layer results in a gradual red-shift of the modes –arising from a gradual increase of its refractive index— and paving the way to a multilevel tunable filtering functionality, rather than a simpler binary configuration. This behaviour is well-reproduced in simulations (shown in 4b), where intermediate crystalline states were calculated using the COMSOL Multiphysics effective medium approach tool (based on the Maxwell-Garnett approximation).[38] By fitting the resonant peak positions of the experimental semicrystalline spectrum, we determined that the partially crystallised experimental state corresponds to approximately 20% of crystallisation fraction in our model. Importantly, the device exhibits absolute modulation depths in transmission (MDT) of up to 60% (Figure 4c): values comparable to those obtained in previous larger-geometry devices. In Figures 4d–e we present the corresponding experimental and calculated reflectance spectra for the same device, which here again were found to be almost complementary to transmittance spectra (i.e. with transmission resonant dips corresponding to light being mostly specularly reflected by the metasurface). Finally, to assess the robustness of our fabrication approach, we performed reproducibility tests on identical devices produced on different days. As discussed in Supplementary Section S4, reflection and transmission spectra were highly consistent from device to device, suggesting reproducibility and reliability of our approach. Additional PCM metasurfaces with varied geometrical parameters and intermediate crystallisation states were also successfully fabricated and tested (Supplementary Section S3), further demonstrating the versatility of our method for realising PCM metasurfaces with customised reconfigurable infrared band-switching and modulation capabilities over large areas.

### 2.3. Dual-band chiral phase-change metasurfaces

To further explore the potential and versatility of our fabrication approach, we fabricated metasurfaces with reconfigurable chiroptical activity based on left- and right-handed arrays of nanoimprinted triskelion motifs. We choose triskelion-based arrays due to their lack of mirror symmetry, which prevents them from being mapped onto their mirror image by any combination of rotations or translations. Triskelia-based metasurfaces have been widely exploited as nanophotonic architectures for (static) giant circular dichroism, chiral second harmonic generation, and chiral light emission to name a few.[39–42] However, to date the combination of chiral metasurfaces with PCMs towards reconfigurable chiroptical activity has remained largely underexplored, with only a few demonstrations reported in the mid-infrared and relying on complicated three-dimensional nanofabrication protocols.[43,44] In contrast, we here demonstrate a simplified architecture compatible with our scalable fabrication strategy, which allows the tailoring of the chiroptical response of our phase-change metasurfaces down to the red part of the visible spectrum, enabling the spectral tuning of its chiroptical activity from the red into the near-infrared. This capability unlocks new opportunities, e.g. polarisation-sensitive reconfigurable imaging, or optical communications. Figure 5a shows the schematics and geometrical parameters of our phase-change chiral nanophotonic architecture. The triskelia motifs have an arm width of 100 nm, a length of 300 nm, and an arm curvature radius of R= 200 nm. The unit cell is arranged in an hexagonal configuration, with the triskelia patterns rotated by 5°, and a lattice constant of $\Lambda$ = 500 nm. Rotation of the triskelia motifs with respect to the lattice was found to improve the circular dichroism, as rigorously discussed in ref. [39]. In analogy to the hole arrays discussed in the previous section, an m-$TiO_2$ residual layer of 400 nm with a triskelion height of 160 nm was chosen. The entire structure is then capped with 20 nm of $Sb_2Se_3$, and protected with 15 nm of $Al_2O_3$. Based on numerical simulations (supplementary information Section S5), a $Sb_2Se_3$ thickness of 20 nm was deemed a good value to achieve a pronounced spectral shift of the chiroptical response between the visible and infrared upon switching the PCM between amorphous and crystalline states. These design choices enables the metasurfaces to operate as a visible-to-infrared chiral switch.

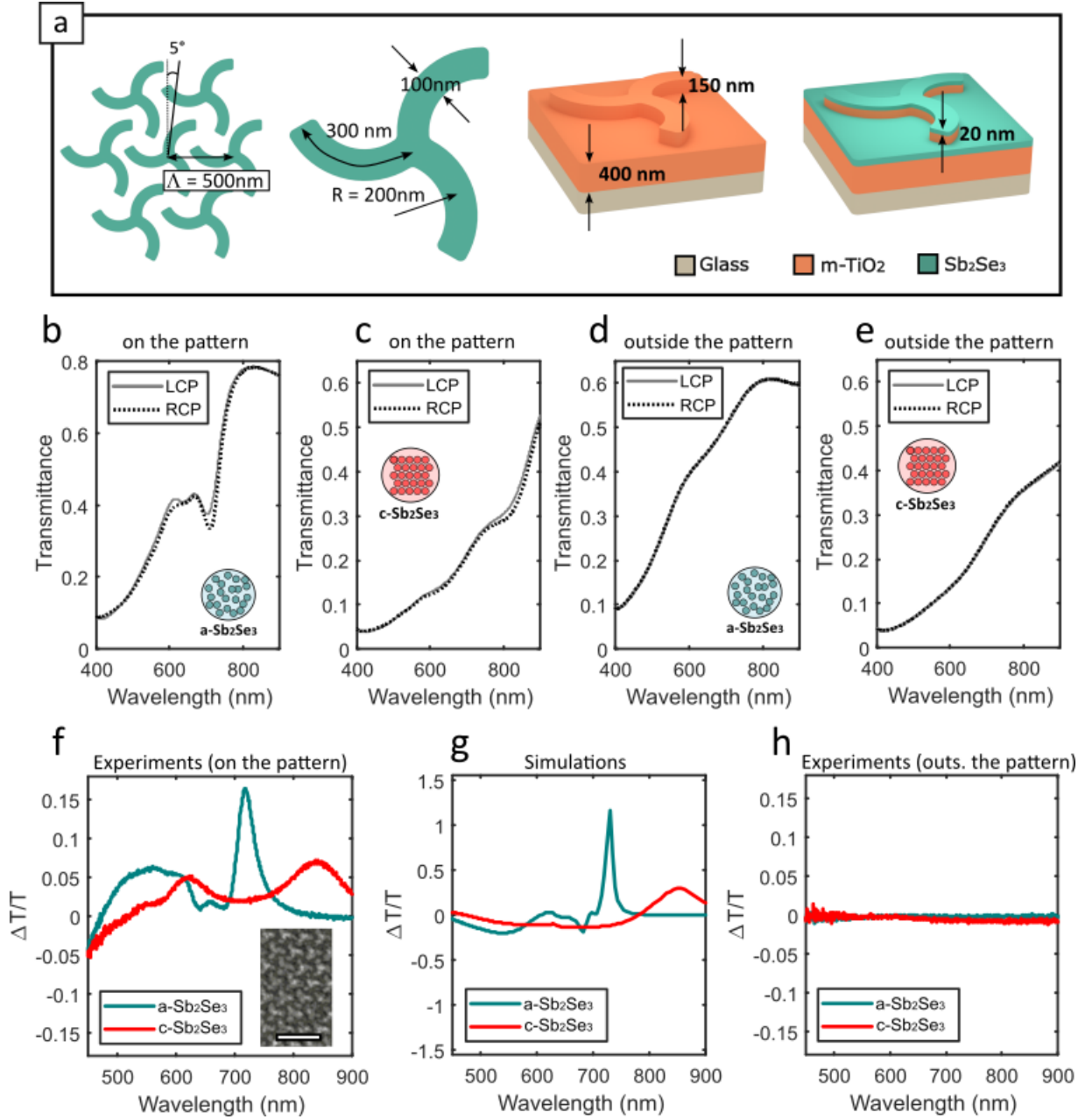

**Figure 5 Reconfigurable chiral metasurfaces based on nanoimprinted m-TiO$_2$ triskelion motifs coated with Sb$_2$Se$_3$.** (a) Schematics of our phase-change chiral metasurface architectures and their geometrical parameters (b–c) Experimental transmittance spectra under left (LCP, grey solid) and right (RCP, black dashed) circularly polarised light for amorphous (b) and crystalline (c) Sb$_2$Se$_3$ states, taken from left-handed triskelion metasurfaces. (d–e) Control transmittance spectra measured outside the patterned area, revealing identical results for both polarisations and Sb$_2$Se$_3$ phases. (f) Experimentally obtained dissymmetry factor (ΔT/T) for amorphous (turquoise) and crystalline (red) states. Inset shows an SEM image of the device (scalebar 1μm). (g) Simulated dissymmetry factor for amorphous and crystalline Sb$_2$Se$_3$ (h) Experimental dissymmetry factor outside the metasurface, confirming the abscence of chiroptical response.

In order to quantify the chiroptical activity of our metasurfaces, we rely on the dissymetry factor *ΔT/T* defined as:[45,46]

$$\frac{\Delta T}{T} = 2 \frac{T_{LCP} - T_{RCP}}{T_{LCP} + T_{RCP}} \qquad \text{Eq. 2}$$

where $T_{LCP}$ and $T_{RCP}$ are the ballistic transmittance spectra for left and right circularly polarised light respectively. In Figures 5b-c, we show the measured transmittance spectra of our left-handed chiral phase-change metasurfaces for LCP (grey solid line) and RCP light (dashed black line), and for amorphous (Figure 5b) and crystalline (Figure 5c) states. These mesurements were carried out at normal incidence in a customised optical setup, whose details are described in detail in section 4.7. Clear differences between LCP and RCP can be already observed in the transmittance spectra when compared to measurements taken outside the patterns (i.e. where transmittance was found to be essentially identical for both polarisations and both states, as revealed by Figures 5d and 5e). In Figure 5f, we show the experimental dissymetry factor, where a resonant dissymetry factor peak of $\Delta T/T \sim 0.15$ can be observed around $\lambda = 715$ nm when the $Sb_2Se_3$ layer is in its amorphous phase (i.e. almost at the red edge of the visible spectrum). After crystallisation, this peak undergoes a remarkable red shift of $\Delta\lambda \sim 120$ nm, relocating it in the near infrared (specifically around $\lambda = 840$ nm). Besides, and in line with previous results from hole-arrays, the crystalline state exhibits a broader peak as a result of resonance damping introduced by the non-negligible absorption coefficient in crystalline $Sb_2Se_3$. Crucially, despite this spectral broadening, the chiral switching functionality remains robust, spanning from the visible to the near-infrared upon crystallisation. This behaviour is further reproduced by our numerical simulations performed in Lumerical (see section 4.4. for simulation details), and displayed in Figure 5g, revealing once again a good qualitative agreement with experimental results. As a comparison, the experimental dissymetry factor outside the samples was also computed ( Figure 5h) where no chiroptical activity was observed for any of the $Sb_2Se_3$ structural phases, further confirming that the dissymetry resonant peaks are coming exclusively from phase-change nanophotonic architectrue presented herein. Finally, analogous experiments were also carried out on right-handed triskelion metasurfaces, with results shown in the supplementary information (Section S6). As expected, the measurements revealed a mirrored spectral behavior, that is, a dual-band chiroptical response upon switching from amorphous to crystalline states, but with opposite signs of the dissymetry factor. This is fully consistent with the enantiomeric nature of the structures, and further demonstrates the visible-to-infrared chiral switching capablities of our scalable phase-change metasurfaces.

## 3. Conclusion

In summary, we have demonstrated a novel, scalable, and thermally robust routine for the fabrication of reconfigurable phase-change metasurfaces, by strategically combining high-throughput direct nanoimprint lithography of m-$TiO_2$ architectures with $Sb_2Se_3$ coatings. Due to the high thermal resilience of m-$TiO_2$, our approach overcomes the limitations of low-melting point polymer-based NIL, paving the way to the integration of phase-change materials into large-area devices capable of withstanding the characteristic high temperatures required for the phase-change transitions. The versatility and resolution of our approach has been illustrated through a set of devices having two different functionalities: infrared band switches and modulators with absolute modulation depths as high as 60%, and chiral metasurfaces with dissymetry factors of up to $\Delta T/T \sim 0.15$, whose chiral properties can be switched between the visible and the near-infrared. To the best of our knowledge, this work also provides the first experimental demonstration of chiral phase-change metasurfaces operating in the visible and near-infrared regimes.[43,44] Overall, the excellent agreement between experiments and simulations, combined with geometrical scalability and reproducibility across multiple devices highlights the robustness of our platform. Finally, it is worth noting that the "all-dielectric" nature of the metasurface architectures reported here may *a priori* seem to limit the practical use of resistive heaters for in-situ switching of the phase-change layer, as recently reported in refs [47,48]. However, in Figure S7 from the supplementary information, we show how imprinting of m-TiO$_2$ can be also readily performed on conductive materials such as metals, thereby opening the door to future cheap and cost efficient PCM-based metasurfaces with in-situ switching capabilities employing metal resistive heaters. Therefore, our approach can be generalised and further explored towards the creation of cheap, and thermally robust phase-change reconfigurable metasurfaces over larger areas, and with exciting functionalities additional to those shown herein, including reconfigurable beam steering devices, image pre-processors, or tunable color displays to name but a few.

## 4. Experimental Section

**4.1. Materials:** Acetone, Isopropanol, Ethanol, and glass substrates were purchased from Labbox. The soft PDMS kit (Sylgard 184) was purchased from Dow Corning (Midland, Michigan, USA). The compounds for synthesizing hard PDMS were acquired from Gelest (USA). Perfluorooctyl-trichlorosilane for the master's silanization was acquired from Merck. Titania paste (90-T) based on 20 nm diameter anatase nanoparticles was purchased from Dyesol (greatcellsolar), Elanora. The Sb$_2$Se$_3$ target for thin film depositions was purchased from Testbourne (Netherlands).

*4.2. Fabrication of h-PDMS/s-PDMS stamps*: Working stamps for imprinting of m-TiO$_2$ were based on hybrid h-PDMS/s-PDMS composites. To prepare PDMS stamps with pillars for the fabrication of hole arrays, the main silicon masters (consisting of arrays of holes) were cleaned in acetone and isopropyl alcohol (10 min ultrasonication each), and subsequently splashed with deionised water followed by drying with compressed air. Next, and in order to prevent permanent adhesion of PDMS to silicon, clean masters were silanized with 12 µl of perfluorooctyl-trichlorosilane, under low vacuum in a desiccator for 30 min. The silanized masters were then post baked for 20 min at 150°C on a hot plate. After the masters were clean and silanized, a s-PDMS 10:1 mixture (monomer:curing agent) was mixed vigorously using a hand stirrer, and left degassing for one hour and a half. Meanwhile, the h-PDMs was prepared by mixing 0.85 g vinylmethylsiloxane, 25 µL of 1,3,5,7-tetracetylcyclosilane, 2 µL of Pt catalyst, 275 µL of hydroxyl siloxane, and 1 mL of toluene under vigorous mixing assisted by a magnetic stirrer. The h-PDMS was then drop cast onto the master, and spread using a compressed air gun. The drop casting and air spreading process was repeated for a couple of times. The substrates were left for 30 min at room temperature followed by an annealing procedure on a hot plate at 60 °C for 1 H. Finally, degassed s-PDMS was poured onto the masters and thermally cured at 100 °C for approximately 1H. For preparing h-PDMS/s-PDMS with holes to make triskelion arrays of pillars, the process was essentially the same, except from the fact that the main working masters were made of Ormostamp pillar replicas obtained from main silicon masters consisting of arrays of triskelion holes.

***4.3. Thin film deposition***: Sb₂Se₃ thin films were prepared by pulsed laser deposition from pure (99.99%) Sb₂Se₃ one inch targets. An ArF excimer laser (λ = nm) was used for the deposition, with a pulse duration of XX ns, a repetition rate of 10 Hz, an energy of 50 mJ and a spot size of xx mm². The Argon pressure inside the vacuum chamber was set to xxx mbar.

***4.4. Numerical calculations***: Finite element calculations for the hole arrays were performed employing the commercial software package COMSOL Multiphysics (RF module), via rigourous solving of Maxwell's equations. To mimic infinite arrays of element, Floquet periodic boundary conditions were applied to the lateral domains of the unit cell shown in Figure 3a, whereas top and bottom domains were truncated using periodic ports. The refractive index of glass substrates and m-TiO₂ were assumed to be constant across the entire IR due to their low dispersion in this spectral regime, with $n_{glass}$=1.5 and $n_{m-TiO2}$= 1.6)[49]. The complex refractive indices of amorphous and crystalline Sb₂Se₃ shown in Figure 1(b) were obtained by ellipsometry measurements. Triskelia arrays were simulated employing the commercial software (FDTD Lumerical from ANSYS). The m-TiO₂ triskelia array had an arm width and height of 100 and 160 nm, respectively. A conformal coating layer of 20 nm was added of the phase change materials with the same optical properties used for the hole arrays. A pair of plane waves was injected with a phase delay of ±90° for LCP and RCP. The ballistic transmittance of both circular polarisations was collected using a power monitor to mimic the experimental characterization for low NA objectives.

***4.5. Phase-change annealing experiments:*** Partial crystallisation of Sb₂Se₃-based metasurfaces was achieved by annealing the devices on a hot plate for 1 min at the material crystallisation temperature (170 ºC)[18]. Full crystallisation was carried out by increasing the annealing time from 1 min to 5 min. Additional crystallisation tests were carried out at higher temperatures (i.e. 190 ºC), but no significant differences on the device optical properties were found. However, and in line with previous studies, annealing at temperatures above 190 ºC and/or longer times were found to severely degrade the metasurface performance. This behavior has been observed previously, and has been attributed to selenium loss and diffusion through the capping layers (here Al₂O₃), which can result in a loss of the optimum stoichiometry. [31,50]

***4.6. Reflectance and Transmittance optical measurements:*** Reflectance and transmittance spectra were acquired through FTIR measurements (BRUKER Vertex 70 equipped with a microscope and a HgCdTe cooled detector). A low numerical aperture objective (NA 0.1) was used to minimize the light cones of excitation and collection, thereby preventing possible resonance broadening and/or generation of additional modes due angular dispersion arising from periodic structures.

For reflection measurements, a reference baseline was taken by collecting the signal from a silver mirror with an average reflectance of R=97%, which was then used to normalize the raw spectra from the samples. Air was used as a reference baseline for transmission measurements. Both types of measurements were carried out with unpolarized light.

*4.7. Circular dichroism measurements in transmission:* Circular dichroism of triskelia-based phase-change metasurfaces was measured employing a home-built optical setup. [46]Briefly, the excitation part was based on a fibre-coupled white lamp (Ocean Optics, HL-2000-HP, FL, USA) with an output reflective collimator made of protected silver (RC08SMA-P01, Thorlabs). A Glan-Thompson linear polarizer (GTH10M, Thorlabs) was placed before the sample in order to obtain linearly polarized light, and an achromatic quarter-wave plate (SAQWP05M-700) was subsequently employed to obtain either right or left circularly polarized light (i.e. by mounting the quarter-wave plate at an angle of ±90° with respect to the axis of the input linearly polarized light). The sample was then placed between two low NA objectives (NA 0.1), here again to reduce the cone of excitation and collection angles and ensure quasi-normal incidence. Finally, light was collected employing a second reflective collimator (protected silver) fiber coupled to a spectrometer (Ocean Optics, QEPro-FL). Transmittance spectra for RCP, LCP were taken by normalizing the raw signal from the sample to the raw signal collected from air for each polarisation state. The dark signal from the spectrometer was subtracted from every raw measurement before normalisation.

*4.8. SEM/EDS measurements:* SEM and EDS measurements were taken with a Quanta 200 ESEM FEG from FEI. Images were acquired under low vacuum (60 Pa) and with an acceleration voltage of 15kV.

**Acknowledgements**
CRdG acknowledges funding from the Marie Sklodowska Curie Individual Fellowship METASCALE (101068089). This work also received funding from the Spanish AEI through grant PID-2022-141956NB-I00 (OUTLIGHT) and the Severo Ochoa Excellence programme CEX2023-001263-S (MATRANS24). This research was also supported by the EIC PATHFINDER CHALLENGES project 101162112 (RADIANT), funded by the European Union. CDW acknowledges funding from the EPSRC (grants EP/W022931/1 and EP/W003341/1).The authors are grateful to Videsh Kumar for providing h-PDMS hole-stamps for the imprinting of triskelia m-TiO$_2$ patterns, and to Afroditi Koutsogianni and Ana Conde-Rubio for sharing time of their SEM sessions. Authors would like to thank as well Dr. Anna Esther Carrillo of the ICMAB electron microscopy Service for her assistance with EDS analysis.

**Data Availability Statement**
The data that support the findings of this study are available from the corresponding author upon reasonable request.

**Supporting Information**

S1. SEM/EDS analysis of the Sb$_2$Se$_3$ films.

S2. Transmission measurements of Glass/m-TiO2/Sb$_2$Se$_3$ unpatterned multilayer stacks.

S3. Optical characterisation of additional phase-change hole arrays for band switching and modulation.

S4. Device reproducibility tests.

S5. Dependence of the dissymmetry factor with the Sb2Se3 layer thickness.

S6. Optical characterisation of the right-triskelion enantiomer.

S7. Fabrication of m-TiO$_2$ nanostructures on different substrates: